\documentclass[doublecol]{epl2}
\usepackage{graphicx}
\usepackage{amsmath}
\usepackage{wasysym}
\usepackage{amssymb}
\usepackage{dcolumn}
\usepackage{bm}
\usepackage{framed}
\usepackage{esint}
\usepackage{epstopdf}
\usepackage{url}

\newcommand{\ket}[1]{\left \vert #1 \right \rangle}

\newcommand{\sigx}{\sigma_x}

\newcommand{\sigz}{\sigma_z}

\DeclareMathOperator{\RRe}{Re}
\DeclareMathOperator{\IIm}{Im}

\newcommand{\sigzeq}{\sigma_{\rm{z,eq}}}
\newcommand{\gonen}{\bar{\gamma}_a}
\newcommand{\gtwon}{\bar{\gamma}_b}
\newcommand{\szeq}{s_{\rm{z,eq}}}
\newcommand{\wm}{\omega_{\rm{m} }}
\newcommand{\gm}{\gamma_{\rm{m} }}
\newcommand{\wmtil}{\tilde{\omega}_{\rm{m} }}
\newcommand{\gmtil}{\tilde{\gamma}_{\rm{m} }}

\newcommand{\Icc}{I_{\rm{cc}}}
\newcommand{\phiezero}{\phi_{\rm{e0}}}
\newcommand{\phieone}{\phi_{\rm{e1}}}
\newcommand{\wdrive}{\omega_{\rm{d}}}

\newcommand{\ORabi}{\Omega_{\rm{R}}}
\newcommand{\gcrit}{g_{\rm{c}}}

\begin{document}

\title{Semiclassical dynamics of a flux qubit coupled to a nanomechanical oscillator}

\date{\today}

\author{Lior Ella \and Eyal Buks}

\institute{Department of Electrical Engineering, Technion, Haifa 32000 Israel}

\abstract{
We present a theory describing the semiclassical dynamics of a superconducting flux qubit inductively coupled to a nanomechanical oscillator. Focusing on the influence of the qubit on the mechanical element, and on the nonlinear phenomena displayed by this device, we show that it exhibits retardation effects and self-excited oscillations. These can be harnessed for the generation of non-classical states of the mechanical oscillator. In addition, we find that this system shares several fundamental properties with cavity optomechanical systems, and elucidate the analogy between these two classes of devices.}


\maketitle

Mechanical devices operating in the quantum limit are currently at the focus of a great deal of research effort~\cite{aspelmeyer_cavity_2013,meystre_short_2012,oconnell_macroscopic_2010}. This is due to their potential for experimentally probing the transition from the microscopic realm, which is described by the rules of quantum mechanics, to the macroscopic one, in which predictions that stem from these rules are almost never directly observed.
In optomechanical cavities, the mechanical element is actuated and detected by coupling it to an electromagnetic resonator, either at microwave~\cite{regal_measuring_2008,teufel_nanomechanical_2009} or optical~\cite{groblacher_demonstration_2009,purdy_observation_2013} frequencies.

In addition to a cavity, the mechanical element can be coupled to an inherently nonlinear device~\cite{bennett_laser-like_2006,schwab_spring_2002,schneider_coupling_2012,etaki_motion_2008,etaki_dc_2011,etaki_self-sustained_2013,buks_decoherence_2006,buks_displacement_2007,blencowe_quantum_2007,rabl_cooling_2010,kolkowitz_coherent_2012} which functions as a two-level system,
such as a flux qubit~\cite{omelyanchouk_quantum_2010,grajcar_sisyphus_2008,hauss_single-qubit_2008}.
The advantage of this scheme is that in contrast to the cavity, which is commonly linear, with this scheme
coherent and periodic energy swaps between the mechanics and the qubit are possible~\cite{oconnell_macroscopic_2010}.
In addition, the energy level splitting of the qubit is highly tunable, and can be varied by controlling the applied flux.
However, despite the differences between the two level system and the linear cavity, the generic nature of the retardation and frequency-shifting effects commonly seen in cavity optomechanics leads to their appearance also in qubit-resonator systems.

In this letter we show this by analyzing the semiclassical dynamics of these two coupled devices.
In particular, we show that when the system is operated in the blue and red-detuned regimes, the change in mechanical resonance frequency and dissipation coefficient is renormalized by terms analogous to the Stokes and anti-Stokes terms~\cite{schliesser_radiation_2006,teufel_circuit_2011,aspelmeyer_cavity_2013} familiar from cavity optomechanics. Next, we focus on the blue-detuned regime of operation, in which self-excited oscillations occur~\cite{zaitsev_forced_2011,etaki_self-sustained_2013,carmon_temporal_2005,armour_quantum_2012,qian_quantum_2011,ludwig_optomechanical_2008}. We analyze the nonlinear dynamics of the device in this regime and show that the mechanical mode can lose its stability as the coupling strength, controlled by an external magnetic field, is increased. We then study the resulting limit cycle dynamics of the device, and derive their amplitude and frequency. A discussion of the physical parameters appropriate for this device are given in~\cite{buks_decoherence_2006}. Note that though we specialize the treatment to a flux qubit, the results contained here are generally applicable to other systems in which a mechanical element is coupled to a two level system.

It has recently been shown that extending the experimental scope of optomechanics to include nonlinear effects provides access to a wide range of new phenomena which can be instrumental for the purpose of driving the mechanical element into a non-classical state~\cite{marquardt_dynamical_2006,ludwig_optomechanical_2008,katz_signatures_2007,carmon_temporal_2005,zaitsev_forced_2011,armour_quantum_2012,qian_quantum_2011}.
In particular, self-excited oscillations, which occur when a driven system loses its stability
and begins to oscillate at one of its resonance frequencies as its control parameters are varied.
This phenomenon, also known as a Hopf bifurcation,
has been studied in the context of cavity optomechanics by several authors~\cite{armour_quantum_2012,ludwig_optomechanical_2008,marquardt_dynamical_2006}, who have demonstrated its potential for detecting signatures of non-classical behavior and enhancing it.

\begin{figure}
\includegraphics[width=1\columnwidth]{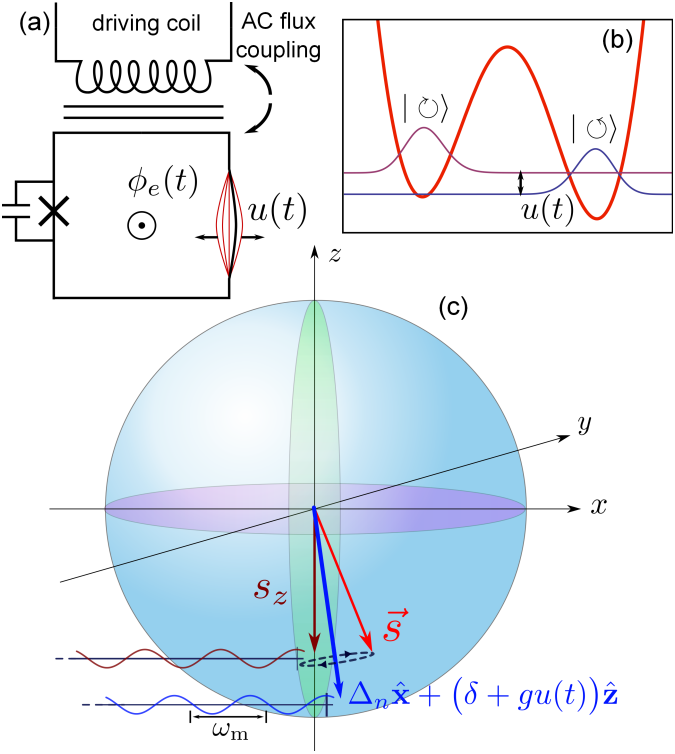}
\caption{\label{fig:schematic}
(a) The rf-SQUID, which is operated as a flux qubit, has a vibrating arm whose position of center of mass $u=\alpha + \alpha^{\ast}$ alters the flux through the SQUID loop. Concurrently, the circulating current in the SQUID, in the presence of a magnetic field, leads to a Lorentz force acting on the beam. (b) The double well potential of the circulating current near the half-flux quantum biasing point of the SQUID. This potential leads to two localized circulating current states $\ket{\circlearrowleft}$ and $\ket{\circlearrowright}$, which span the qubit's Hilbert space. (c) An illustration of dynamics of the qubit coupled to the mechanical element in the semiclassical picture and in the rotating frame, for the linear regime. Here a sinusoidal oscillation of the beam leads to a response of the z-component of the qubit, which is proportional to the circulating current. This response, given in~(\ref{eq:qubit_response_func}) and~(\ref{eq:stokes_anti_stokes}), leads to a renormalization of the mechanical dissipation coefficient and resonance frequency.}
\end{figure}

In what follows we analyze the dynamics of a flux qubit with a vibrating arm that functions as a mechanical oscillator~\cite{buks_decoherence_2006,buks_decoherence_2012,buks_displacement_2007} (see Fig.~\ref{fig:schematic}). We assume that this system is described by the Hamiltonian~\cite{buks_decoherence_2006,wal_quantum_2000,berns_coherent_2006,hauss_single-qubit_2008}
\begin{equation}
H=H_{\rm S} + H_{\rm R} + H_{\rm SR} \label{eq:H_general}
\end{equation}
where
\begin{align}
H_{\rm S}  &=\frac{1}{2}\varepsilon_{0}\phi_e(t)\sigz+\frac{1}{2}\hbar\Delta\sigx\nonumber
  +\hbar\wm a^{\dagger}a \\
  &+\frac{1}{2}\hbar g(a+a^{\dagger})\sigz. \label{eq:H_S_def}
\end{align}
The first two terms in~(\ref{eq:H_S_def}) account for the effective two level system, whose states correspond to localized circulating current $\ket{\circlearrowleft}$ and $\ket{\circlearrowright}$ in the SQUID loop (see Fig.~\ref{fig:schematic}).  
Here $\sigma_x$ and $\sigma_z$ are Pauli matrices, $a$ and $a^\dagger$ are the lowering and raising operators of the mechanical mode. the applied magnetic flux
$\phi_e(t) = \phiezero + \phieone\cos(\wdrive t)$
has a DC component corresponding to the difference in energy of the two circulating current states and an AC component originating from the externally applied microwave signal, and
$\hbar \Delta$ is the energy difference at the degeneracy point $\phiezero=0$. We assume that the qubit is engineered and biased such that $\varepsilon_0\phiezero \gg \hbar \Delta$ and is strongly driven, with $\varepsilon_0\phieone/\hbar \simeq \wdrive$. In this case, and when $k_B T \ll \hbar \Delta$, the state at thermal equilibrium is a localized circulating current state, and transitions to the excited state proceed via multi-photon resonances~\cite{berns_coherent_2006}.
The third term in $H$ is related to the mechanical mode, where $\wm$ is its resonance angular frequency.

The fourth term represents the coupling between the qubit and the mechanical mode, and results from the dependence of the flux threading the qubit loop on the amplitude of the mechanical mode. Here $\hbar g=Bl\Icc\sqrt{\hbar/2m\wm}$, where $B$ is the applied magnetic field, $l$ is the effective length of the suspended beam, $I_{cc}$ is the magnitude of the localized circulating current, and $m$ is the effective mass of the mechanical mode.

The Hamiltonians $H_{\rm R}$ and $H_{\rm SR}$ correspond, respectively, to the Hamiltonian of bosonic reservoirs representing the environment and to the interaction terms of the mechanical mode and the qubit with the environment. We assume that the modes of the reservoirs couple to $a+a^\dagger$, $\sigma_x$, and $\sigma_z$, which introduces mechanical relaxation, qubit thermal relaxation and qubit dephasing, respectively.

To eliminate the explicit time dependence in \eqref{eq:H_S_def}, we transform to the interaction picture with
$U(t) = \exp\left(-i\frac{1}{2}\sigma_z \varepsilon_0 \int^t \phi_e (t')\rm{d}t'\right)$.
Assuming that $\varepsilon_{0}\phiezero\simeq n\hbar\wdrive$
for a particular integer $n$ this leads to the Hamiltonian in the rotating frame
\begin{align}
H_{\rm S}  &=\frac{1}{2}\delta \sigz+\frac{1}{2}\hbar\Delta_n\sigx\nonumber
  +\hbar\wm a^{\dagger}a \\
  &+\frac{1}{2}\hbar g(a+a^{\dagger})\sigz. \label{eq:H_S_interaction}
\end{align}
where $\delta=\varepsilon_{0}\phiezero/\hbar-n\wdrive$
and
$\Delta_{n}=\Delta J_{n}\left(\varepsilon_{0}\phieone/\hbar\wdrive\right).$
Here $J_n$ is the Bessel function of the first kind~\cite{shevchenko_multiphoton_2012,shevchenko_resonant_2008,shevchenko_landauzenerstuckelberg_2010}, and a rotating wave approximation (RWA) has been performed.

Since we wish to focus on how the response of the qubit influences the dynamics of the mechanical mode, 
it is sufficient to consider the semiclassical equation
formed by averaging these equations over the degrees
of freedom of the qubit and the oscillator. We note
that when $g=0$, the dynamics of the qubit and the oscillator is
independent. This implies that the covariance function between the
oscillator and the qubit coordinates is of order $g$, which allows
us to neglect it when $g$ is small. With this approximation all correlations of the system operators factorize, and~(\ref{eq:H_general}) leads to the following equations of motion:
\begin{subequations}
\label{eq:EOM_semiclassical}
\begin{align}
\dot{s}_{-}(t) & =-\gamma_{2}s_{-}(t) - i\delta s_{-}(t) + i\frac{1}{2}\Delta_{n}s_{z}(t)\label{subeq:EOM_semiclassical_sm}\\
 &-ig\bigl(\alpha(t)+\alpha^{\ast}(t)\bigr)s_{-}(t),\nonumber \\
\dot{s}_{z}(t) & =-\gamma_{1}\left(s_{z}(t)-\sigma_{z,{\rm eq}}\right)\label{subeq:EOM_semiclassical_sz}\\
 & +i\Delta_{n}\bigl(s_{-}(t)-s_{+}(t)\bigr),\nonumber \\
\dot{\alpha}(t) & =-i\wm\alpha(t)-\frac{\gm }{2}\alpha(t)-i\frac{1}{2}g s_{z}(t),\label{subeq:EOM_semiclassical_a}
\end{align}
\end{subequations}
where $\szeq$ is the thermal equilibrium value of $s_z$,
\begin{equation}
s_{\pm}=\frac{1}{2}\left\langle\sigma_x \pm i\sigma_y\right\rangle,\, s_{z}=\left\langle \sigma_{z}\right\rangle,\,\alpha=\left\langle a\right\rangle,
\end{equation}
and the averaging is over the degrees of freedom of the reservoir and the system, which we have assumed to be statistically independent.

The interpretation of~(\ref{eq:EOM_semiclassical}) is straightforward: The qubit evolution is described by the Bloch equations, and the mechanical amplitude is that of a harmonic oscillator. The coupling is manifested as a force on the mechanical mode, 
on one hand, and as a change in the detuning frequency of the qubit on the other.

We now determine how the mechanical mode is influenced by the coupling to the qubit. We first discuss the renormalization of mechanical dissipation coefficient and resonance frequency, and then focus on self-excited oscillations of the system in the blue-detuned regime.

\begin{figure}
\includegraphics[width=1\columnwidth]{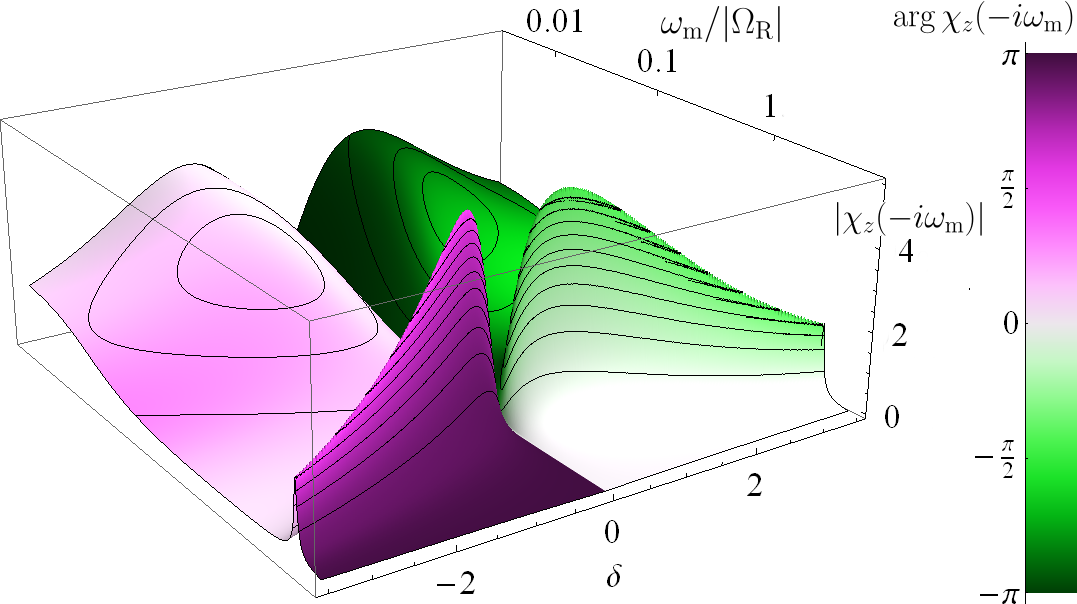}
\caption{\label{fig:qubit_response_full}The response function $\chi_z(-i\wm)$ as given in~\eqref{eq:qubit_response_func},
as a function of $\delta$ and $\omega$. The vertical axis corresponds
to the absolute value, the color to the phase, and the
contours to the imaginary part. For all $\delta<0$, the
phase is positive, which implies that the qubit adds delay, thus
decreasing the effective dissipation coefficient of the oscillator. For $\delta>0$, the
opposite is true. Note that here, in addition to the resonant peak, another smaller peak appears at $\wm=\gonen$ (see also Fig.~\ref{fig:Real_and_Im})}
\end{figure}

\begin{figure}
\includegraphics[width=1\columnwidth]{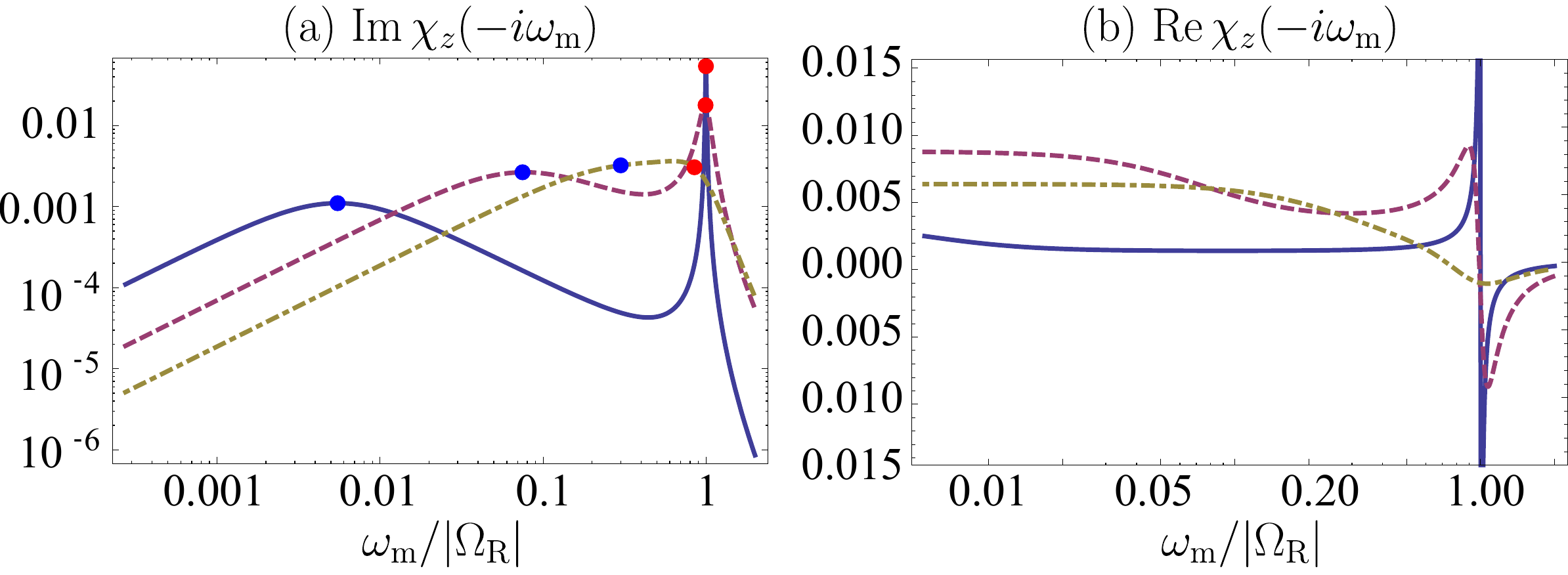}
\caption{\label{fig:Real_and_Im}The imaginary (a) and real (b) parts of the qubit response function, $\chi_{z}(-i\wm)$, as given by~\eqref{eq:qubit_response_func} in the blue-detuned ($\delta<0$) regime, for typical system parameters and different
qubit decay times $\gamma_1$ and $\gamma_2$. The blue dots in (a) corresponds to the maximum at $\omega = \gonen$, and the red dots to the maximum at $\omega = \bar{\Omega}_{\rm R}$. Solid line: $\gamma_1=0.001$, $\gamma_2=0.01$. Dashed line: $\gamma_1=0.05$. $\gamma_2=0.1$, Dot-dashed line: $\gamma_1=0.1$, $\gamma_2=0.5$.}
\end{figure}

We assume that both the dissipation and the coupling
are small, namely that $\gamma_{1}\simeq\gamma_{2}\ll\delta\simeq\Delta_{n}$,
$\gm \ll\wm$, and $g\ll\wm$, and that $\wm$
is never significantly larger than 
\begin{equation}
\ORabi ={\rm sign}(\delta)\sqrt{\delta^{2}+\Delta_{n}^{2}},
\end{equation}
the Rabi frequency of the qubit, defined here to have the same sign as $\delta$.
We consider small deviations from the equilibrium point of the qubit-oscillator system found by setting the time derivatives in~(\ref{eq:EOM_semiclassical}) to zero. In the linear regime, a periodic oscillation of the mechanical amplitude $\alpha(t) = \alpha_{\rm eq} + \alpha_0 e^{-i\wm t}$ around its equilibrium position will lead to a response $s_z(t) = s_{z,\rm eq} + \chi_z(-i\wm)\alpha_0 e^{-i\wm t}$ (see Fig.~\ref{fig:schematic}). This response, when fed back to the mechanical amplitude equation~(\ref{subeq:EOM_semiclassical_a}), will lead to a renormalization of $\gm$ and $\wm$~\cite{ella_nonlinear_2014}:
\begin{align}
\gmtil &= \gm -g\IIm\chi_z(-i\wm),\label{eq:gmre_general}\\
\wmtil &= \wm+\frac{1}{2}g\RRe\chi_z(-i\wm),\nonumber 
\end{align}
where we find that this response function is given by
\begin{equation}
\begin{gathered}
\chi_{z}(-i\wm)= \\
\frac{-2\ORabi G \bigl(2\gamma_{2}-i\wm\bigr)}{\bigl(\gonen-i\wm\bigr)\bigl(\gtwon-i(\wm-\ORabi)\bigr)\bigl(\gtwon-i(\wm+\ORabi)\bigr)},\label{eq:qubit_response_func}
\end{gathered}
\end{equation}
and is plotted in Fig.~\ref{fig:qubit_response_full}. Here
$G = -\delta\gamma_1\Delta_n^2 \sigzeq g/2\ORabi^3\gonen$
is a (positive) interaction coefficient with a magnitude dependent on the equilibrium point of the qubit,
and
\begin{align}
\gonen &= \frac{\delta^2\gamma_1 + \Delta_n^2\gamma_2}{\ORabi^2},\nonumber \\
\gtwon &= \gamma_{2}-\frac{\Delta_{n}^{2}}{2\ORabi ^{2}}(\gamma_2 - \gamma_1).\label{eq:gamma_approx_def}
\end{align}
The location of the poles and zeros of~(\ref{eq:qubit_response_func}) is correct to second order in the small parameters as specified above.
The real and imaginary parts of $\chi_z(-i\wm)$ are plotted in
Fig.~\ref{fig:Real_and_Im}, and a plot of $\chi_{z}$
as a function of $\omega$ and $\delta$ is given in Fig.~\ref{fig:qubit_response_full}.

To make contact with cavity optomechanics, we note that when the qubit and the mechanical element are resonant, with $|\ORabi|\simeq\wm$, and the coherence times of the qubit are long, the qubit-oscillator interaction takes on a form very similar to the Stokes and anti-Stokes terms
\begin{subequations}
\label{eq:stokes_anti_stokes}
\begin{align}
\IIm\chi_z(-i\wm) &=
G\left(\frac{\gtwon}{\gtwon^2 + (\wm + \ORabi)^2}
- \frac{\gtwon}{\gtwon^2 + (\wm - \ORabi)^2}\right)  \label{eq:SAS_Im}\\
\RRe \chi_z(-i\wm) &=
G\left( \frac{\wm + \ORabi}{\gtwon^2 + (\wm + \ORabi)^2} + \frac{\wm - \ORabi}{\gtwon^2 + (\wm - \ORabi)^2}\right) \label{eq:SAS_Re}
\end{align}
\end{subequations}
We see that $\gtwon$ plays the role of the cavity damping rate, and $\Omega_{\rm R}$, whose sign is equal to the sign of the detuning $\delta$, corresponds to the cavity detuning. In contrast to cavity optomechanics, however, here the response exhibits a richer structure with an additional peak at $\wm=\gonen$, as can be seen in Figs.~\ref{fig:qubit_response_full} and~\ref{fig:Real_and_Im}.

\begin{figure}
\includegraphics[width=1\columnwidth]{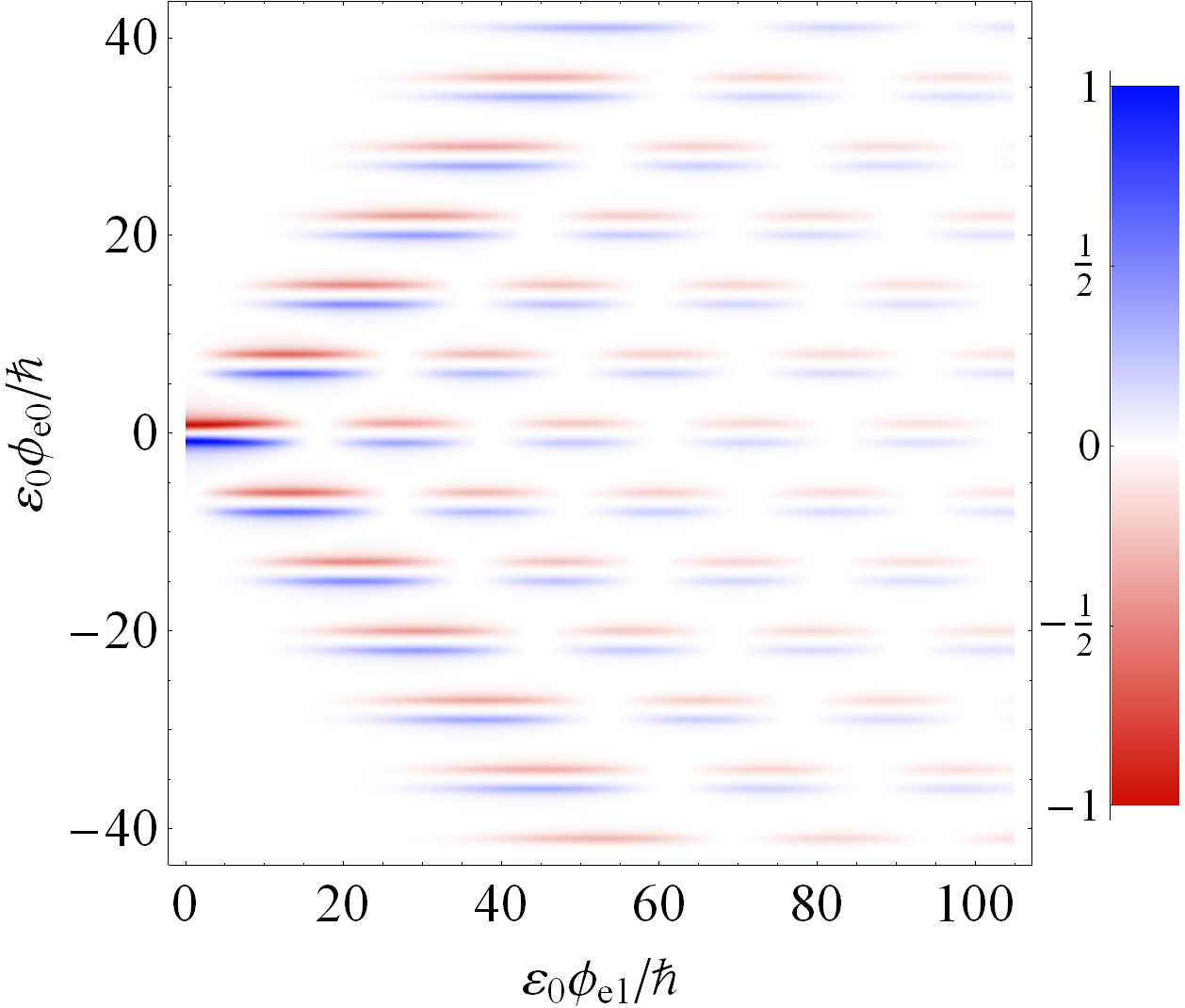}
\caption{\label{fig:Correction-to-dissipation}Correction of dissipation coefficient
as a function of external flux driving amplitude, $\varepsilon_{0}\phieone/\hbar$,
and detuning, $\varepsilon_{0}\phiezero/\hbar$. The blue areas corresponds
to a decrease, and the red to an increase, in the effective mechanical
dissipation coefficient. The color intensities are scaled by the extremal values of the correction. This plot corresponds to the resonance regime, with $\wm = 1.28\Delta$. 
The values of the correction are given in arbitrary units. Parameters used are: $\gamma_{1}=0.014$, $\gamma_{2}=0.714$,
$g=0.0018$, $\Delta=0.1$, all scaled to $\wdrive = 1$. $\sigma_{z,{\rm eq}}=-1$, and $Q=10^{5}$. This plot may be compared to Refs.~\cite{berns_coherent_2006,shevchenko_landauzenerstuckelberg_2010}.}
\end{figure}

The general behavior of the correction to the mechanical dissipation
coefficient, as a function of the DC and AC parts of the externally
applied flux, $\varepsilon_{0}\phiezero$ and $\varepsilon_{0}\phieone$, can
be seen in Fig.~\ref{fig:Correction-to-dissipation}. This result
was obtained by superposing the corrections for different values of
$n$, the multi-photon Rabi resonance.
This result may be compared to the Landau-Zener interference diagrams,
given in Refs.~\cite{berns_coherent_2006,shevchenko_landauzenerstuckelberg_2010}.

Next, we study the self-excited oscillations that the system exhibits in the case $\Omega_{\rm R}<0$,  when the system is blue-detuned. For this we no longer neglect the nonlinear part of~(\ref{eq:EOM_semiclassical}).
We find analytical expressions for the amplitude and frequency of the limit cycle of both the oscillator and the qubit, in the regime where the system is approximately
resonant, i.e.~that $\wm\simeq|\ORabi| $. This is done by diagonalizing the non-interacting part of~(\ref{eq:EOM_semiclassical}) and performing a RWA~\cite{ella_nonlinear_2014}. We find that for $\gmtil >0$
the system will have a single stable equilibrium at the origin, while
for $\gmtil <0$, the origin will lose its stability and
a limit cycle will emerge. For the original equations~(\ref{eq:EOM_semiclassical}), this corresponds
to a supercritical Hopf bifurcation.

Solving for the amplitude of the limit cycle, we find that the critical coupling for which $\gmtil=0$ and a limit cycle appears is given by
\begin{equation}
\gcrit =\sqrt{\frac{2\gm \ORabi^2 \left(\gtwon{}^{2}+\sigma^{2}\right)}{\szeq\gtwon\Delta_{n}^{2}}}. \label{eq:g_crit}
\end{equation}
Furthermore, we find that when $g\gtrsim \gcrit$ the amplitude of the limit cycle is given by
\begin{align}
r_{s} &= \frac{\szeq}{2}\sqrt{\frac{\gonen}{\gtwon}}\frac{\gcrit ^{2}}{g^{2}}\sqrt{\frac{g^{2}}{\gcrit ^{2}}-1},\nonumber \\
r_{a} &= \sqrt{\frac{\gonen\szeq}{2\gm }}\frac{\gcrit }{|g|}\sqrt{\frac{g^{2}}{\gcrit ^{2}}-1},\label{eq:nonzero_FP}\\
s_{cz} &= \szeq\left(\frac{\gcrit ^{2}}{g^{2}}-1\right),\nonumber \\
\omega_{a} &= \frac{\gm \sigma}{2\gtwon+\gm },\nonumber \\
f(\sigma) &= \arctan\frac{2\sigma}{2\gtwon+\gm }.\nonumber 
\end{align}
Here $\sigma = \wm - |\ORabi|$ and $\szeq=\delta \gamma_1 \sigzeq/|\ORabi|\gonen$.
The amplitudes are defined as $\bar{\alpha}_{c}=r_{a}\exp\left(i\omega_{a}t\right)$, $\bar{s}_{c-} =r_{s}\exp\left[i\left(\omega_{a}t-\frac{\pi}{2}+f(\sigma)\right)\right]$,  and overbar denotes that they are in the diagonal basis of~(\ref{eq:EOM_semiclassical}).

\begin{figure}
\includegraphics[width=1\columnwidth]{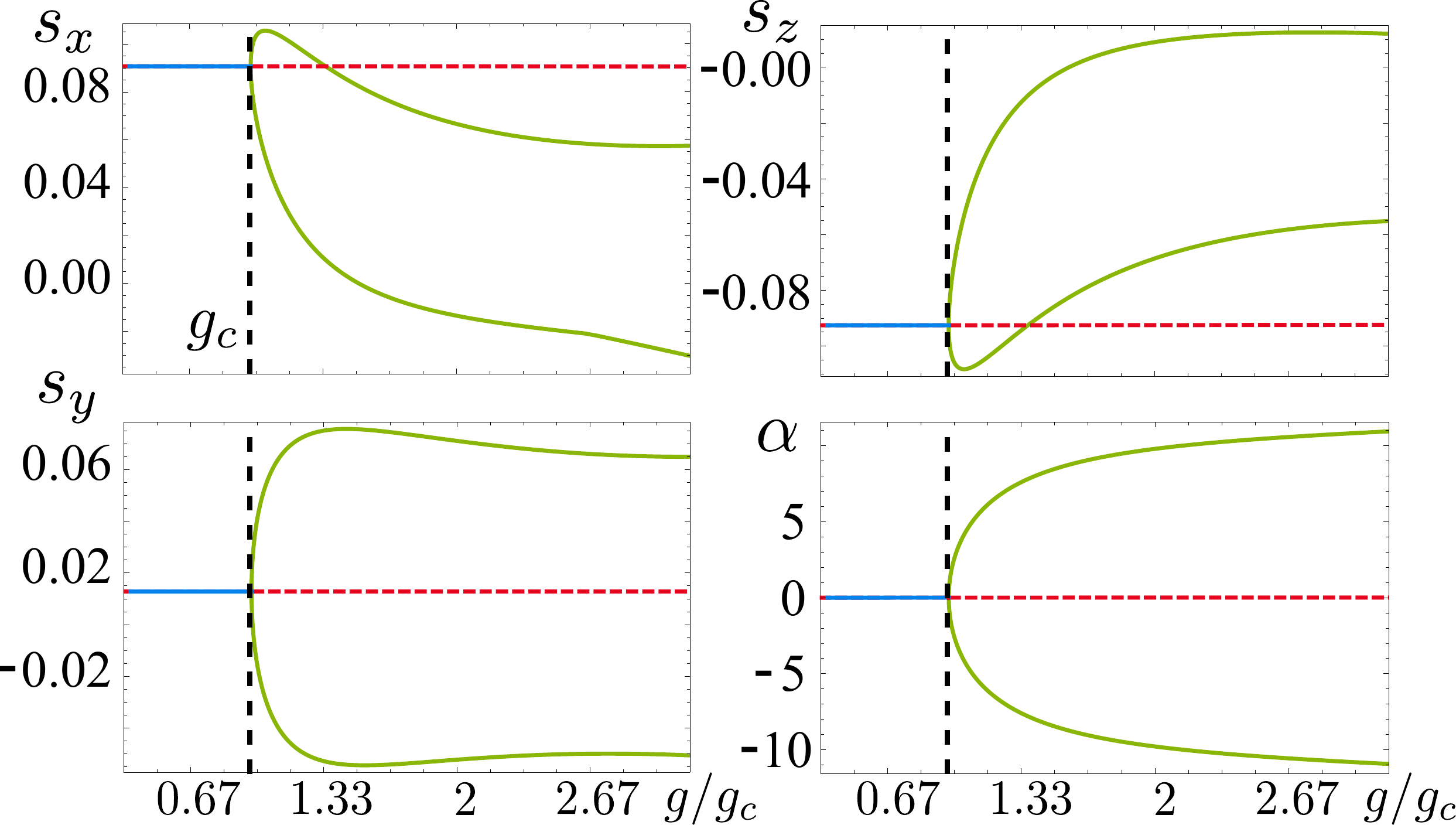}
\caption{\label{fig:Bifurcation-curves-for}
near resonance, with $\wm=1.1\ORabi $. The blue and dashed
red lines correspond to stable and unstable equilibrium points,
respectively. The green curve corresponds to a stable limit cycle,
where the top curve corresponds to the maximal value of the variable
during the limit cycle, and the bottom curve corresponds to the minimal
value.}
\end{figure}

Thus the system has a limit cycle for $g>\gcrit $,
with amplitudes proportional to $\sqrt{g-\gcrit }$, as is the case
with non-degenerate supercritical Hopf bifurcations. Furthermore,
the frequency of the limit cycle solution undergoes a shift proportional
to $\sigma$, as can be seen in the equation for $\omega_{a}$ in~(\ref{eq:nonzero_FP}).
In Fig.~\ref{fig:Bifurcation-curves-for} we can see the bifurcation
curves for the system near resonance, which were calculated using
numerical continuation.

We have found that when described using a semiclassical approximation, the dynamics of a flux qubit coupled to a nanomechanical oscillator
via a coupling of the form $(\hbar g/2)(a+a^{\dagger})\sigz$ influences the mechanics
in a manner similar to that found in cavity optomechanics: In the linear regime, we have shown that the mechanical dissipation coefficient and resonance frequency
are renormalized with expressions given in~(\ref{eq:gmre_general}), (\ref{eq:qubit_response_func}) and~(\ref{eq:stokes_anti_stokes}).
These expressions reveal that the condition for resonance in this system is that the mechanical frequency equals the Rabi frequency of the qubit,
and that the resolved sideband limit is at $\wm\gg\gtwon$, where $\gtwon$ is defined in~(\ref{eq:gamma_approx_def}). They also show
that in this case the response has a richer structure, with an additional peak at $\wm = \gonen$.
Considering the possibility of multi-photon driving of the qubit, we have shown that the Stokes and anti-Stokes sidebands
of the qubit response exhibit a Bessel-ladder behavior~\cite{berns_coherent_2006,shevchenko_landauzenerstuckelberg_2010},
as shown in Fig.~\ref{fig:Correction-to-dissipation}. 

Extending our analysis to the nonlinear regime for a blue-detuned qubit,
we have shown that the system exhibits self-excited oscillations when the coupling $g$,
whose strength is controlled by an external magnetic field, is increased beyond $g_c$ which is given in~(\ref{eq:g_crit}).
We have found the amplitude of the limit cycle close to criticality in~(\ref{eq:nonzero_FP}),
and calculated numerically its behavior for general $g$,
as shown in Fig.~\ref{fig:Bifurcation-curves-for}.  

The limit cycle behavior of this system suggests a possible scheme for the preparation of
non-classical entangled states of the qubit and oscillator. Since the limit cycle dynamics
can be described by a single phase variable, if the system is cooled to its ground state
so that thermal noise is negligible and then rapidly brought to a limit cycle state, this
phase variable can be expected to be found in a superposition state. A more quantitative
analysis of this point is left to subsequent articles.

This work was supported by the German Israel Foundation under Grant No.~1-2038.1114.07, the
Israel Science Foundation under Grant No.~1380021 and the European STREP QNEMS Project.

\bibliographystyle{eplbib}
\bibliography{thesis_bibliography}

\end{document}